\pdfoutput=1
\documentclass[11pt]{article}

\usepackage[a4paper,margin=2.4cm]{geometry}
\usepackage[T1]{fontenc}
\usepackage{mathptmx}
\usepackage[scaled=0.9]{helvet}
\usepackage{courier}
\usepackage{microtype}
\usepackage{amsmath}
\usepackage{graphicx}
\usepackage{array}
\usepackage{booktabs}
\usepackage{makecell}
\usepackage{multirow}
\usepackage[table]{xcolor}
\usepackage[font=small,labelfont=bf,labelsep=period]{caption}
\usepackage[numbers,sort&compress]{natbib}
\usepackage{enumitem}
\usepackage[section]{placeins}
\usepackage{titlesec}
\usepackage{hyperref}
\hypersetup{colorlinks=true,linkcolor=cblue!70!black,citecolor=cblue!70!black,urlcolor=cblue!70!black,pdftitle={Report-Derived Labels, an Image Teacher and the Gap to Expert Labels in Knee MRI},pdfauthor={Soheil Kooklan}}

\definecolor{cblue}{HTML}{2A78D6}
\definecolor{corange}{HTML}{EB6834}
\definecolor{caqua}{HTML}{1BAF7A}
\definecolor{cgrey}{HTML}{E6E5E0}
\definecolor{cink}{HTML}{52514E}

\titleformat{\section}{\normalfont\large\bfseries}{\thesection}{0.7em}{}
\titleformat{\subsection}{\normalfont\normalsize\bfseries}{\thesubsection}{0.7em}{}
\titlespacing*{\section}{0pt}{2.2ex plus .5ex}{1ex}
\titlespacing*{\subsection}{0pt}{1.6ex plus .4ex}{0.6ex}
\setlist{itemsep=2pt,topsep=4pt}

\title{\bfseries Report-Derived Labels, an Image Teacher and the Gap to Expert Labels in Knee MRI:\\[2pt] An Analysis of One Entry to the RSNA 2026 Knee Abnormality Detection Challenge}
\author{Soheil Kooklan\\[2pt]
\normalsize Independent Researcher\\
\normalsize \texttt{soheilkooklan@gmail.com}\\
\normalsize ORCID \href{https://orcid.org/0009-0003-5035-7833}{0009-0003-5035-7833}}
\date{28 October 2026}

\begin{document}
\maketitle

\begin{abstract}
\noindent
In the RSNA 2026 Knee Abnormality Detection challenge, 58 of 4{,}407 training studies have expert labels for 12 findings and the rest only a radiology report. We compare training targets built from a language-model reading of the reports and from a public image model, against two references: the expert labels and the 39{,}248 study-finding cells that the reports state as present or absent. The report labels and the out-of-fold predictions of the image model reach similar agreement with the expert labels (macro AUC 0.892 and 0.896) but fail on different findings. The report is better for ligament tears and patellofemoral osteoarthritis, which it never calls present when the experts do not; the image model is better for effusion, Baker's cyst, contusion and synovitis, where the reports call the finding absent, leave it out or call it present more often than the experts. Their equal mixture reaches 0.927 (gain over the report labels 0.035, 95\% interval 0.019 to 0.052), also after removing the studies whose knee side was handled inconsistently. We did not observe a model-level gain from this better target: a 2.5D ResNet50 trained on the report labels alone scored 0.877 on the private leaderboard, after the deadline, against 0.872 for the same model trained on the mixture, and the out-of-fold advantage of the mixture is confounded by a change of code and by possible leakage through the image model's folds. A scanner prevalence prior raised agreement with the reports but not with the experts, so a report-derived reference can reward reporting habits. A rank correction for co-occurring findings gave small out-of-fold gains and no visible leaderboard effect. Our entry, a forked public ensemble with that correction, was displayed at rank 582 of 6{,}299 on the private leaderboard (0.943, a score shared by 287 teams). All results are exploratory; none of the models is meant for clinical use.

\end{abstract}

\section{Introduction}
\label{sec:intro}

Image models for knee MRI are usually trained with labels taken from radiology reports, because expert reading of every study is too expensive~\cite{bien2018mrnet,qiu2024coplane}. The report is written for a referring clinician, though, and it leaves out findings that did not matter for the question asked, grades some findings by a local habit, and names others that are hard to see on the images. A model trained on report labels learns all of this. How far report-derived labels are from expert labels, finding by finding, is usually not measured, because the expert labels are missing.

The RSNA 2026 Knee Abnormality Detection challenge~\cite{rsna2026news,kaggle2026knee} has both, in very unequal amounts. Its training set has 4{,}407 knee MRI studies with their DICOM series and reports, in at least nine languages. For 58 of them, two musculoskeletal radiologists and an adjudicator set labels for 12 findings: anterior cruciate ligament (ACL) tear, medial collateral ligament (MCL) tear, medial and lateral meniscus tear, medial, lateral and patellofemoral (PF) osteoarthritis (OA), joint effusion, synovitis, Baker's cyst, bone contusion and fracture. Test studies have neither reports nor labels. Entries are notebooks that score each test study offline within nine hours, ranked by the mean of the 12 per-finding ROC AUCs, on a public part of the test set during the challenge and on a private part of about 70\% at the end.

\begin{figure}[!t]
\centering
\includegraphics{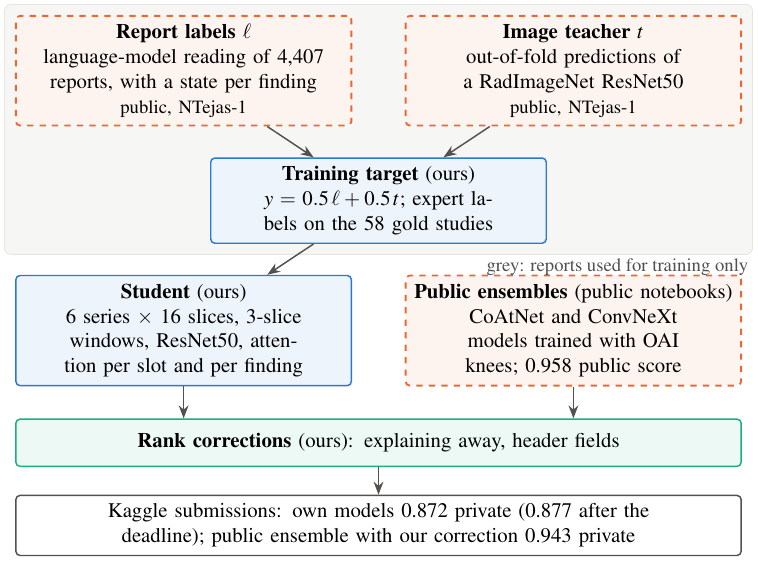}
\caption{What this report analyses. Orange with dashed border: material made by others; blue and green with solid border: our work. Reports enter only the training target; test studies have no report.}
\label{fig:overview}
\end{figure}

This report analyses one entry. We built a training target from a public language-model reading of the reports and a public image model, trained our own 2.5D model on it, and in the last days of the challenge tried post-hoc corrections on public ensembles. Figure~\ref{fig:overview} separates what we built from what we took from others. The 58 expert-labelled studies are too few to rank close variants, so we report every comparison against two references with different strengths, and say where they disagree.

The contributions are empirical:
\begin{enumerate}[label=(\arabic*)]
\item a per-finding comparison of report labels, image-model predictions and their mixtures as training targets against the expert labels, with paired intervals for both contrasts and a count of the ways in which the reports disagree with the experts;
\item a 2.5D student trained on the mixed target and, with the same recipe, on the report labels alone, compared out of fold and on the hidden test, where no gain from the better target was observed;
\item an example in which a correction improves agreement with the reports and not with the expert labels, which shows a limit of report-derived references for model selection;
\item a set of post-hoc rank corrections, including one based on co-occurring findings, reported with the full parameter grid, a correction for multiple testing and their leaderboard effect.
\end{enumerate}
Our final rank came from a public ensemble that we forked, and our own models scored 0.872 and, after the deadline, 0.877 on the private leaderboard against 0.943 for that ensemble with our correction. We do not claim a competitive system, and none of the models is intended or validated for clinical use.

\section{Related work}
\label{sec:related}

\paragraph{Knee MRI models.} MRNet~\cite{bien2018mrnet} classified ACL tears, meniscal tears and abnormal knees from three series per study with labels taken from reports. \citet{qiu2024coplane} diagnosed the same twelve kinds of knee abnormality as this challenge with attention across planes and sequences. fastMRI+~\cite{zhao2021fastmriplus} added bounding-box pathology labels to the fastMRI knee data. Our model belongs to the multiple-instance family: instances are 3-slice windows, pooled by gated attention~\cite{ilse2018attention} within a series and by one attention query per finding across series, with a RadImageNet encoder~\cite{mei2022radimagenet}.

\paragraph{Labels from reports.} CheXpert~\cite{irvin2019chexpert} labelled chest radiographs with a rule-based reader that keeps an uncertain state, and CheXbert~\cite{smit2020chexbert} trained a language model on such labels plus expert annotations. Large language models are now used to extract structured findings from reports, with accuracy that depends on the model and the prompt~\cite{dhanaliwala2023llm}. These studies evaluate the labeller against expert reading of the report. Here the reference is expert reading of the images, so report omissions count as errors.

\paragraph{Soft targets and co-occurrence.} Training on the predictions of another model is knowledge distillation~\cite{hinton2015distilling}; mixing a teacher with noisy labels is one of its simplest forms. In multi-label recognition, label co-occurrence is usually built into the model, for example with a graph over labels~\cite{chen2019mlgcn}. We use it in the opposite direction, after training, to remove some of the dependence between scores; the name comes from explaining away in Bayesian networks~\cite{wellman1993explaining}, although our correction is a linear heuristic with no probabilistic model behind it.

\section{Data and validation references}
\label{sec:data}

\subsection{Report statements}
\label{sec:explicit}

\begin{figure}[!t]
\centering
\includegraphics[width=0.62\linewidth]{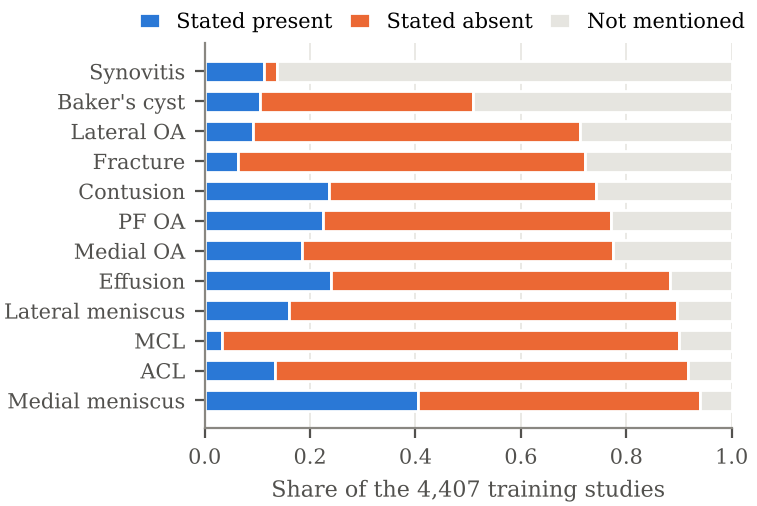}
\caption{What the 4{,}407 training reports say about each finding, according to the report labels. The explicit reference uses only the stated cells.}
\label{fig:coverage}
\end{figure}

The report labels we use (Section~\ref{sec:targets}) give every finding of every study a state: stated present, stated absent or not mentioned. Of the 52{,}884 study-finding cells, 39{,}248 are stated (74.2\%), 8{,}754 of them as present. Figure~\ref{fig:coverage} shows how this differs by finding. Synovitis is stated in 14\% of the studies (604 cells), Baker's cyst in 51\%, the medial meniscus in 94\%. We call the AUC against the stated cells the explicit AUC. With about 880 studies per fold it can separate variants, and it does not depend on any image model. It measures agreement with what the radiologist wrote, as read by a language model, which is a different thing from what the challenge scores.

\subsection{Expert labels}

The second reference is the expert labels of the 58 gold studies. According to the challenge material, they were set by two musculoskeletal radiologists with an adjudicator; whether the readers saw the reports is not stated in the material we had. Their positive counts are: ACL 24, MCL 9, medial meniscus 26, lateral meniscus 23, medial OA 15, lateral OA 11, PF OA 21, effusion 35, synovitis 27, Baker's cyst 12, contusion 19 and fracture 18. With 9 to 35 positives, the standard error of an AUC near 0.85 is about 0.04 to 0.08~\cite{hanley1982auc}. Every gold result in this report is therefore exploratory, and we treat a variant as better only when both references move in the same direction.

\subsection{Folds and the teacher's folds}
\label{sec:folds}

All our models use one partition into five folds of 881 or 882 studies (10 to 12 gold studies each), with byte-identical reports kept in one fold (seed 2026). The released tables have no patient identifier. The DICOM headers give a distinct PatientID for each of the 4{,}407 studies, but these may have been assigned per study at anonymisation, so two studies of one patient could still fall into different folds, and we do not call the split patient-independent. Gold studies are trained on their expert labels at weight 2, and each gold study is scored only by the model of the fold that held it out. The fold manifest is released with the code.

The image teacher (Section~\ref{sec:targets}) was trained by its authors with their own five folds, which we do not have. Its prediction for each study is out of fold in that partition, so the target-agreement analysis of Section~\ref{sec:targetres} is not affected. Our student is affected: when fold $k$ is held out, the targets of the other studies include teacher predictions from teacher models that may have been trained on studies of fold $k$, and in this way the report labels and images of fold $k$ can reach the student's training targets. We could not rebuild the teacher with our folds, so the student's out-of-fold scores in Section~\ref{sec:student} may be optimistic by an amount we cannot size. We read them as upper estimates and draw no inferential conclusion from the difference between student and teacher. The leaderboard scores are free of this path, and so is a student trained on the report labels alone (Section~\ref{sec:labonly}).

\subsection{Statistics}
\label{sec:stats}

For the difference between two score sets on the gold studies, we report the observed difference in macro AUC on the 58 studies and a 95\% percentile interval from 1{,}000 bootstrap resamples of the studies (seed 0). A finding without positives or negatives in a resample is skipped in that resample. Where a $p$ value is given, it comes from a paired permutation test in the manner of \citet{bandos2005permutation}, which we chose because it needs no normal approximation, and with 9 to 35 positives per finding a normal approximation is doubtful. Both score sets are turned into ranks within each finding, so that they share one scale; the two sets are swapped within a study, for all twelve findings at once, with probability one half; and the macro AUC difference is recomputed 1{,}000 times. The $p$ value is the share of permuted differences at least as large in absolute value as the observed one, counting the observed one. The null hypothesis of this test is that the two score sets are exchangeable within each study. That is a stronger statement than equal macro AUC, so a small $p$ says that the two sets rank the studies differently in a way that favours one of them, and the test should not be read as a general test of equal macro AUC. As a check that does not assume exchangeability, though it does use a normal approximation, we also computed a paired test of DeLong type~\cite{delong1988}: each study contributes its DeLong placement values for all twelve findings, so the dependence between findings scored on the same studies enters the variance, and for a single finding the test is the usual paired DeLong test. Holm adjustment~\cite{holm1979} is applied within one family of tests, the 15 explaining-away settings of Section~\ref{sec:eares}; no adjustment is made across the other comparisons of this report, which is one more reason to read them as exploratory. All numbers in Sections~\ref{sec:targetres} to~\ref{sec:refres} are produced by one script from the released files (Section~\ref{sec:availability}).

\section{Methods}
\label{sec:methods}

\subsection{Training targets}
\label{sec:targets}

We use two public files from the NTejas-1 repository~\cite{ntejas2026repo} (MIT licence). The first holds a reading of all 4{,}407 reports by Claude Opus 4.5, with a probability and a state per finding; we call these the report labels $\ell$. The second holds out-of-fold predictions $t$ of an image model from the same repository, a RadImageNet ResNet50 at 224\,px trained on those report labels; we call it the teacher. The target of study $i$ and finding $j$ is
\begin{equation}
y_{ij} = (1-w_j)\,\ell_{ij} + w_j\,t_{ij}, \qquad w_j = 0.5 .
\end{equation}
The share 0.5 was set before we scored any target against the gold studies; we kept no dated record of that choice, so this comparison, like the others, is exploratory. If one source is missing for a cell, the other is used, and if both are missing the target is 0.1; neither case occurs in the released files. The target agreement AUC of a candidate target is its AUC against the expert labels. It describes the target, not a trained model.

\subsection{First entry}

Our first entry used a multilingual rule-based reader of the reports (target agreement 0.776), an EfficientNet-B3 over 3-slice instances, and a training schedule that left most of the encoder frozen; its augmentation function was never called. It scored 0.735 public and 0.713 private. Everything below belongs to the second entry.

\subsection{Image cache}
\label{sec:cache}

Every study is decoded once into an 8-bit array of shape [6 series slots, 16 slices, 256, 256]. The slots are the sagittal, coronal and axial planes, each split by the challenge's fluid-sensitive flag; the series with the most files fills a slot, and an empty slot stays empty and is masked (80.7\% of the slots are filled, no study is empty). Within a series, the slices are sorted by \texttt{ImagePositionPatient} along the patient axis with the largest spread. For parallel slices this order is either the order along the stack normal or its exact reverse, whenever the normal has a component on that axis; a unit test checks this for 200 random orientations. The axis order always runs from the lower to the higher patient coordinate, so it points in the same anatomical direction for every series of one plane, while the direction of the stack normal depends on the sign of the orientation vectors in the header. The direction matters, because the model reads adjacent slices as ordered channels. Series without enough positions fall back to the instance number. Sixteen slices are taken evenly over the 6 to 94\% band of the stack, each is cropped to 140\,mm using the row pixel spacing, scaled to its own 1st and 99th percentiles and resized to 256\,px. This per-slice window, kept from the NTejas-1 format, removes gain differences between scanners and series, and it also removes real intensity differences between neighbouring slices. A slice that fails to decode is replaced by another slice of the same series. Laterality comes from the DICOM tag, or from the median $x$ of \texttt{ImagePositionPatient} when it is at least 20\,mm from zero, and right knees are mirrored to a left-knee convention. This is the image format of the NTejas-1 models, and the decode functions are copied from that repository.

\subsection{Student model and training}
\label{sec:model}

An instance is a window of three consecutive cached slices used as three image channels. A RadImageNet ResNet50~\cite{mei2022radimagenet,he2016resnet} (inputs in $[-1,1]$, weights from a public Kaggle copy~\cite{marwan2026radimagenet}) encodes each window. Gated attention~\cite{ilse2018attention} pools the windows of a slot into one vector, and each finding has its own attention query over the six slot vectors, with absent slots masked. A learnable $12\times6$ bias is added to these attention logits, initialised from the series a radiologist reads first for each finding (Appendix Table~\ref{tab:prior}); it is trained with the rest of the head.

The released out-of-fold predictions come from checkpoints trained by notebook~04 with the settings below and the 224\,px crop. Training draws 4 random windows per slot, a random 224\,px crop of the 256\,px slices, and a gain in $[0.9,1.1]$ and offset in $[-8,8]$ grey levels per study. There are no flips, because laterality is fixed in the cache. Validation and test use all 14 windows of a slot and the centre crop. The whole encoder is trained with AdamW~\cite{loshchilov2019adamw} for 8 epochs (learning rate $2\cdot10^{-4}$ for the encoder and $2\cdot10^{-3}$ for the head, weight decay 0.01, linear warm-up over 8\% of the steps, then cosine decay), two studies per step with gradient accumulation over four, mixed precision and gradient clipping at 3, against the soft targets with binary cross-entropy. The weights of the last three epochs are averaged~\cite{izmailov2018swa}. One fold takes about 68 minutes on two T4 GPUs. At test time the five fold models are averaged in probability space, and model families are combined in rank space. Each fold was trained once.

\subsection{Rank corrections}
\label{sec:ea}

Several findings of the knee occur together: OA of a compartment with a tear of its meniscus, a bone contusion with an ACL tear or a fracture. A model that learns these associations raises the score of a meniscus in a knee with OA whether or not the meniscus is torn. Let $r_{ij}$ be the percentile rank of study $i$ for finding $j$ over the scored set, and let $W$ be the positive part of the correlation matrix of the report labels over the 4{,}349 non-gold studies, with a zero diagonal and rows summing to 1. We use
\begin{equation}
r'_{ij} = r_{ij} + \lambda \sum_k W_{jk}\, r_{ik}
\label{eq:ea}
\end{equation}
for every finding except effusion and synovitis, which are left unchanged. With $\lambda<0$ a finding gives up rank when its usual partners rank high (we call this explaining away, by analogy only); with $\lambda>0$ the partners support it. The largest weights are fracture to contusion (0.30), medial meniscus to medial OA (0.23), PF OA to medial OA (0.23), ACL to contusion (0.22) and lateral meniscus to lateral OA (0.22).

We also read one DICOM header per study. Patient age, weight, birth date, study date and institution are empty in every study. Sex is present in 94.6\% of training studies, the manufacturer and model in all (12 manufacturer strings, 46 models), and the field strength in 94.6\% (2{,}543 studies at 1.5\,T, 1{,}601 at 3\,T). From these we built three kinds of correction: a per-finding weight on sex, chosen on one random half of the studies and applied to the other half and the reverse; subtracting a share of each scanner group's mean rank, computed on the scored set itself as it would be on a test set; and adding a scanner prevalence prior, the report-label rate of each scanner group in one half, shrunk towards the overall rate with a pseudo-count of 20, applied to the other half.

\section{Results}
\label{sec:results}

Table~\ref{tab:evidence} lists the kinds of evidence in this report and what each can show. The sections below follow the same order.

\begin{table}[!t]
\caption{The kinds of evidence used in this report.}
\label{tab:evidence}
\small
\begin{tabular}{>{\raggedright\arraybackslash}p{2.9cm}>{\raggedright\arraybackslash}p{3.5cm}>{\raggedright\arraybackslash}p{3.4cm}>{\raggedright\arraybackslash}p{4.6cm}}
\toprule
Evidence & Reference & Size & What it can show \\
\midrule
Target agreement & expert labels & 58 studies & how close a training target is to the experts \\
Student, out of fold & expert labels; report statements & 58 studies; 39{,}248 cells & how a trained model ranks; gold optimistic (Section~\ref{sec:folds}) \\
Fold-0 runs & both, out of fold & 882 studies; 11 gold & the effect of one design change; three seeds of one configuration \\
Rank corrections & both, out of fold & as above & whether a correction moves both references \\
Leaderboard & hidden test labels & public part; about 70\% private & test score, three decimals only \\
\bottomrule
\end{tabular}
\end{table}

\subsection{Report labels and the image teacher fail on different findings}
\label{sec:targetres}

\begin{figure}[!t]
\centering
\includegraphics[width=0.66\linewidth]{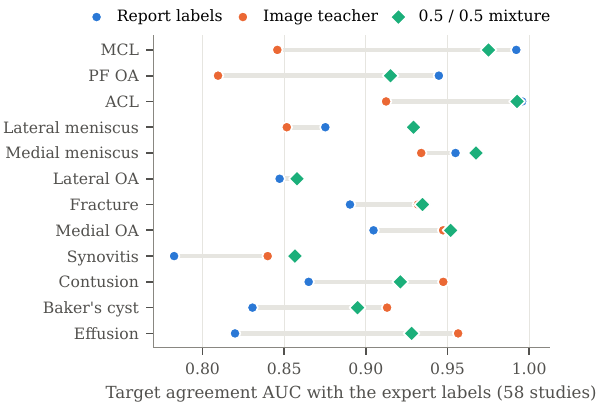}
\caption{Target agreement with the expert labels per finding. Findings are ordered by the difference between the report labels and the teacher. The grey bar spans the two sources.}
\label{fig:complementarity}
\end{figure}

Figure~\ref{fig:complementarity} and Table~\ref{tab:targets} compare candidate targets on the gold studies. The report labels and the teacher reach almost the same macro target agreement (0.892 and 0.896; difference 0.004, interval $-0.026$ to $0.033$), but not on the same findings. The report is far better on findings a radiologist states outright: ACL 0.996, MCL 0.992 and PF OA 0.945, against 0.912, 0.846 and 0.809 for the teacher. The teacher is better on effusion (0.957 against 0.820), contusion (0.947 against 0.865), Baker's cyst (0.913 against 0.831) and synovitis (0.840 against 0.783), which are the findings that reports often leave unmentioned (Figure~\ref{fig:coverage}) or that the annotators judged on the images by a different threshold. The equal mixture reaches 0.927, 0.035 above the report labels (0.019 to 0.052) and 0.032 above the teacher (0.013 to 0.051).

\begin{table}[!t]
\caption{Target agreement AUC of candidate training targets with the expert labels (58 studies). Silent only: the teacher only where the report does not mention the finding. Last rows: macro differences with 95\% bootstrap intervals.}
\label{tab:targets}
\small
\setlength{\tabcolsep}{6pt}
\begin{tabular}{lrrcccc}
\toprule
Finding & Pos. & Neg. & Report labels & Teacher & 0.5 / 0.5 mixture & Silent only \\
\midrule
ACL & 24 & 34 & 0.996 & 0.912 & 0.993 & 0.996 \\
MCL & 9 & 49 & 0.992 & 0.846 & 0.975 & 0.988 \\
Medial meniscus & 26 & 32 & 0.955 & 0.934 & 0.968 & 0.959 \\
Lateral meniscus & 23 & 35 & 0.875 & 0.852 & 0.929 & 0.879 \\
Medial OA & 15 & 43 & 0.905 & 0.947 & 0.952 & 0.889 \\
Lateral OA & 11 & 47 & 0.847 & 0.857 & 0.858 & 0.797 \\
PF OA & 21 & 37 & 0.945 & 0.809 & 0.915 & 0.893 \\
Effusion & 35 & 23 & 0.820 & 0.957 & 0.928 & 0.820 \\
Synovitis & 27 & 31 & 0.783 & 0.840 & 0.857 & 0.800 \\
Baker's cyst & 12 & 46 & 0.831 & 0.913 & 0.895 & 0.848 \\
Contusion & 19 & 39 & 0.865 & 0.947 & 0.921 & 0.864 \\
Fracture & 18 & 40 & 0.890 & 0.932 & 0.935 & 0.899 \\
\midrule
Macro & & & 0.892 & 0.896 & 0.927 & 0.886 \\
\multicolumn{3}{l}{Difference to report labels} & -- & $+0.004$ & $+0.035$ & $-0.006$ \\
\multicolumn{3}{l}{\quad 95\% interval} & & \footnotesize($-0.026$, $0.033$) & \footnotesize($0.019$, $0.052$) & \footnotesize($-0.019$, $0.007$) \\
\multicolumn{3}{l}{Difference to teacher} & & -- & $+0.032$ & \\
\multicolumn{3}{l}{\quad 95\% interval} & & & \footnotesize($0.013$, $0.051$) & \\
\bottomrule
\end{tabular}
\end{table}

Using the teacher only where the report is silent gave 0.886, slightly below the report labels; the authors of the teacher describe the same outcome in their repository~\cite{ntejas2026repo}. Mixing helps on stated cells as well, because a finding the report states is not always the finding the annotators saw. Shares that depend on the report state did not beat the flat mixture: over 0.3 to 0.6 for stated cells and 0.5 to 0.9 for silent ones, the best of 20 settings was the flat one, and the others reached 0.920 to 0.926. A per-finding mixture that gives the report more weight for ACL, MCL and PF OA and the teacher more for effusion, Baker's cyst, contusion and fracture reached 0.935, but we chose its shares after seeing Table~\ref{tab:targets}, so it is optimistic and we did not train with it.

Table~\ref{tab:states} shows where the report labels lose on the gold studies. A report can disagree with the experts in three ways: it states a finding absent that the experts saw, it does not mention a finding the experts saw, or it states a finding present that the experts did not label. For the ligaments all three are rare. The report states an ACL tear present in 23 of the 24 expert-positive studies and in none of the 34 expert-negative ones; for the MCL the counts are 7 of 9 and none of 49. PF OA looks worse in the table than in its AUC. The report states it present in only 10 of 21 expert-positive studies, but it never states it present in an expert-negative study, and the language-model reading gives the 11 missed studies intermediate probabilities (median 0.18, against 0.07 for the expert-negative studies), so the report labels still rank them above the expert-negative studies in 89\% of pairs, against 70\% for the teacher. Effusion shows the first kind of disagreement most clearly: 13 of the 35 expert-positive studies have a report that states no effusion, and the report labels rank the 17 positives they miss above the negatives in 67\% of pairs, the teacher in 93\%. For synovitis the report is silent in 14 of 27 expert-positive studies. For contusion it errs the other way and states a contusion present in 15 of 39 expert-negative studies; this would fit reports that describe marrow oedema which the reading maps to a contusion, but we did not check the report texts. For Baker's cyst, 3 of the 12 expert-positive reports state that there is none.

\begin{table}[!t]
\caption{What the report states on the 58 gold studies, by expert label. Present, absent: the report states the finding present or absent; silent: the report does not mention it.}
\label{tab:states}
\small
\setlength{\tabcolsep}{6pt}
\begin{tabular}{lcccccc}
\toprule
& \multicolumn{3}{c}{Expert positive} & \multicolumn{3}{c}{Expert negative} \\
\cmidrule(lr){2-4}\cmidrule(lr){5-7}
Finding & Present & Absent & Silent & Present & Absent & Silent \\
\midrule
ACL & 23 & 1 & 0 & 0 & 30 & 4 \\
MCL & 7 & 2 & 0 & 0 & 42 & 7 \\
Medial meniscus & 24 & 1 & 1 & 5 & 25 & 2 \\
Lateral meniscus & 19 & 3 & 1 & 5 & 23 & 7 \\
Medial OA & 11 & 4 & 0 & 2 & 27 & 14 \\
Lateral OA & 5 & 5 & 1 & 5 & 21 & 21 \\
PF OA & 10 & 6 & 5 & 0 & 27 & 10 \\
Effusion & 18 & 13 & 4 & 2 & 19 & 2 \\
Synovitis & 11 & 2 & 14 & 3 & 1 & 27 \\
Baker's cyst & 8 & 3 & 1 & 2 & 15 & 29 \\
Contusion & 17 & 1 & 1 & 15 & 15 & 9 \\
Fracture & 13 & 3 & 2 & 5 & 21 & 14 \\
\bottomrule
\end{tabular}
\end{table}

\begin{figure}[!t]
\centering
\includegraphics[width=0.58\linewidth]{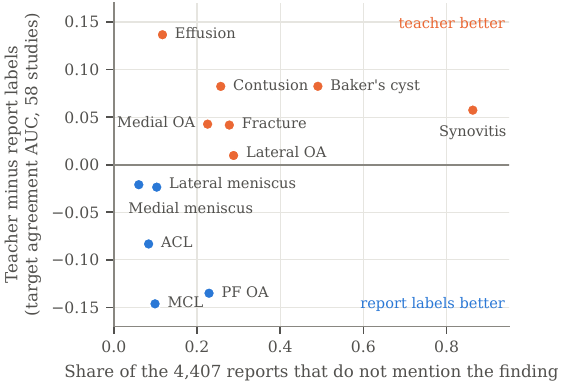}
\caption{Share of the 4{,}407 reports that do not mention a finding (Figure~\ref{fig:coverage}) against the difference in target agreement AUC between the teacher and the report labels on the 58 gold studies (Table~\ref{tab:targets}). Spearman correlation 0.54, permutation $p$ 0.08.}
\label{fig:omission}
\end{figure}

Omission alone explains only part of this (Figure~\ref{fig:omission}). Across the 12 findings, the teacher's advantage tends to rise with the share of reports that leave the finding out, but the association is weak (Spearman 0.54, permutation $p$ 0.08; 0.34 when only the 58 gold reports are counted). Effusion and PF OA break the pattern. Effusion is mentioned in 88\% of the reports and has the largest teacher advantage, because a report that mentions it often calls it absent; PF OA is left out of 23\% of the reports and still favours the report. For effusion, then, the data point to a different threshold between the reporting radiologist and the expert readers more than to omission.

The teacher was built with the same decode functions as our cache and so shares the side handling described in Section~\ref{sec:limits}: in 3 of the 58 gold studies the series of the knee were given opposite sides, and in 9 others at least one series has no side. Removing these studies leaves the result as it was (Table~\ref{tab:side}). On the 46 studies with a consistent side, the mixture is 0.041 above the report labels and 0.036 above the teacher, with intervals that exclude zero, although the smallest positive count falls to 7. The target-level result is therefore a statement about this public teacher with its preprocessing, and it holds on the studies where that preprocessing is consistent.

\begin{table}[!t]
\caption{Target agreement on subsets of the gold studies without the side errors of Section~\ref{sec:limits}. Conflict: series of one study assigned to opposite sides. Unresolved: at least one series with neither a side tag nor a median position at least 20\,mm from zero. Differences with 95\% bootstrap intervals.}
\label{tab:side}
\small
\setlength{\tabcolsep}{4pt}
\begin{tabular}{lrccccc}
\toprule
Gold studies & $n$ & Report & Teacher & Mixture & Mixture $-$ report & Mixture $-$ teacher \\
\midrule
All & 58 & 0.892 & 0.896 & 0.927 & $+0.035$ (0.019, 0.052) & $+0.032$ (0.013, 0.051) \\
No side conflict & 55 & 0.892 & 0.897 & 0.927 & $+0.035$ (0.019, 0.052) & $+0.030$ (0.011, 0.051) \\
No conflict or unresolved side & 46 & 0.882 & 0.887 & 0.923 & $+0.041$ (0.024, 0.057) & $+0.036$ (0.012, 0.061) \\
\bottomrule
\end{tabular}
\end{table}

\subsection{The student stays below its teacher}
\label{sec:student}

Pooled over the five folds, the student reaches 0.876 on the expert labels and 0.863 on the report statements, against 0.896 and 0.879 for the teacher (Table~\ref{tab:oof}). The gold difference is $-0.019$ ($-0.039$ to $-0.002$; DeLong-type $p$ 0.04). The leakage path of Section~\ref{sec:folds} would, if anything, make the student look better, so it does not explain this gap, but it does mean that 0.876 is an upper estimate. Per fold, the explicit AUC was 0.848 to 0.873 and still rising or flat at the last epoch, and the gold AUC of the 10 to 12 gold studies per fold ranged from 0.820 to 0.928 (Appendix Figure~\ref{fig:epochs}). The student is better than the teacher on medial OA, effusion, MCL and lateral OA, and clearly worse on both menisci (0.856 against 0.934, 0.779 against 0.852) and on synovitis. The two models have a mean rank correlation of 0.845 across findings, and their rank mean reaches 0.890, below the teacher alone ($-0.005$, $-0.015$ to $0.004$). Average precision tells the same story (macro 0.828 against 0.831; Appendix Table~\ref{tab:secondary}).

\begin{table}[!t]
\caption{Pooled out-of-fold AUC of the student and of the teacher against the expert labels (58 studies) and the report statements (all studies). The student's gold scores may be optimistic because of the teacher's folds (Section~\ref{sec:folds}).}
\label{tab:oof}
\small
\setlength{\tabcolsep}{8pt}
\begin{tabular}{lcccc}
\toprule
& \multicolumn{2}{c}{Expert labels} & \multicolumn{2}{c}{Report statements} \\
\cmidrule(lr){2-3}\cmidrule(lr){4-5}
Finding & Student & Teacher & Student & Teacher \\
\midrule
ACL & 0.869 & 0.912 & 0.871 & 0.894 \\
MCL & 0.868 & 0.846 & 0.877 & 0.874 \\
Medial meniscus & 0.856 & 0.934 & 0.895 & 0.915 \\
Lateral meniscus & 0.779 & 0.852 & 0.802 & 0.867 \\
Medial OA & 0.964 & 0.947 & 0.907 & 0.908 \\
Lateral OA & 0.865 & 0.857 & 0.837 & 0.849 \\
PF OA & 0.793 & 0.809 & 0.824 & 0.850 \\
Effusion & 0.975 & 0.957 & 0.909 & 0.898 \\
Synovitis & 0.802 & 0.840 & 0.779 & 0.800 \\
Baker's cyst & 0.879 & 0.913 & 0.906 & 0.913 \\
Contusion & 0.945 & 0.947 & 0.858 & 0.878 \\
Fracture & 0.919 & 0.932 & 0.887 & 0.903 \\
\midrule
Macro & 0.876 & 0.896 & 0.863 & 0.879 \\
\bottomrule
\end{tabular}
\end{table}

A student trained on targets that are half teacher would be expected to come close to the teacher. The fold-0 runs of Table~\ref{tab:ablation} test some of the reasons why it did not. All were scored on the same 882 held-out studies, 11 of them gold; differences in the explicit AUC carry a bootstrap interval over these studies, and the expert-label scores of a single fold are listed for completeness only.

The seed matters little on the report reference and more on the 11 gold studies: three seeds of the released code reached explicit AUCs of 0.856, 0.857 and 0.857 and gold AUCs of 0.905, 0.923 and 0.909, and two runs with the same seed gave identical predictions, so the released training is deterministic on this hardware. The original run of the same configuration, made with an earlier version of the code and the Kaggle image of that time, reached 0.861 and 0.928, which is 0.005 above the seed spread on the report reference (interval 0.000 to 0.010 against run 3). We did not isolate which change of code or environment caused this. The slot prior did not help: without it the explicit AUC was 0.860, above all three seeds with it (difference to run 3 0.004, $-0.001$ to 0.010). The earlier cache (288\,px, 12 slices, 130\,mm, no prior, original code) reached 0.849, 0.012 below the original run on the new cache and 0.007 below the seeded runs.

\begin{table}[!t]
\caption{Single runs on fold 0 (882 held-out studies, 11 of them gold). Code: the original training code of the reported runs or the released code with a seeded sampler. Difference: explicit AUC minus that of run 3, with a 95\% bootstrap interval over the 882 studies. Run 4 repeats run 3 and gave identical predictions. The earlier-cache run held out the same studies and five more, which are not scored here.}
\label{tab:ablation}
\small
\setlength{\tabcolsep}{4.5pt}
\begin{tabular}{rllllccc}
\toprule
Run & Cache & Prior & Target & Code, seed & Explicit & Difference (interval) & Gold \\
\midrule
1 & earlier, 288\,px & no & mixture & original & 0.849 & $-0.007$ ($-0.014$, $0.000$) & 0.895 \\
2 & 256\,px, 16 slices & yes & mixture & original & 0.861 & $+0.005$ ($0.000$, $0.010$) & 0.928 \\
3 & 256\,px, 16 slices & yes & mixture & released, 2026 & 0.856 & -- & 0.905 \\
4 & 256\,px, 16 slices & yes & mixture & released, 2026 & 0.856 & $0.000$ & 0.905 \\
5 & 256\,px, 16 slices & yes & mixture & released, 2027 & 0.857 & $+0.001$ ($-0.004$, $0.006$) & 0.923 \\
6 & 256\,px, 16 slices & yes & mixture & released, 2028 & 0.857 & $+0.001$ ($-0.004$, $0.006$) & 0.909 \\
7 & 256\,px, 16 slices & no & mixture & released, 2026 & 0.860 & $+0.004$ ($-0.001$, $0.010$) & 0.924 \\
8 & 256\,px, 16 slices & yes & report labels & released, 2026 & 0.856 & $+0.001$ ($-0.005$, $0.006$) & 0.902 \\
\midrule
& \multicolumn{4}{l}{Teacher (public, single slices, other folds)} & 0.871 & $+0.015$ ($0.006$, $0.024$) & 0.921 \\
\bottomrule
\end{tabular}
\end{table}

The meniscus gap may come from the window. The teacher reads single slices, while our window of three cached slices spans a median of 11.6\,mm of a sagittal series, about four acquired slices, because the cache keeps 16 slices spread over 88\% of the stack (measured on the headers of all training series); thin meniscal tears may be blurred across such a window. We did not run the single-slice variant that would test this. A ConvNeXt-Tiny encoder was also tried and ran at 5.8 to 6.2\,s per step against 0.22\,s for the ResNet50, with no time spent waiting for data, so it was dropped.

We planned a second round in which the student's predictions join the teacher pool. The resulting target reached 0.923 on the gold studies against 0.927 for the first-round target ($-0.004$, $-0.012$ to $0.003$), so the round was not run.

\subsection{A better target did not give a better model in this study}
\label{sec:labonly}

To see whether the better target gives a better model, we trained the same student with the same folds and settings on the report labels alone, with no teacher in its targets (Table~\ref{tab:labonly}). Pooled over the five folds, the mixed target looks better: 0.876 against 0.857 on the expert labels (difference 0.019, interval 0.003 to 0.036, permutation $p$ 0.006, DeLong-type $p$ 0.010) and 0.863 against 0.859 on the report statements (0.004, 0.001 to 0.006), the second on a reference built from the very labels the labels-only student was trained on. On the hidden test the order turned round: the labels-only student scored 0.891 public and 0.877 private, the mixed student 0.887 and 0.872. The labels-only student was submitted after the deadline, so its score is a post-deadline evaluation on the same private test set and not an entry of the challenge.

The pooled comparison is not matched. The mixed student was trained with the original code and the labels-only student with the released code, and on fold~0 that change alone moved the explicit AUC by 0.005 and the gold AUC by 0.023 (Table~\ref{tab:ablation}, runs 2 and 3). With the released code on fold~0, the labels-only student (run 8) lies inside the spread of three seeds of the mixed student on the report reference (0.856 against 0.856 to 0.857) and just below it on the 11 gold studies (0.902 against 0.905 to 0.923). The rest of the pooled gap may come from the leakage of Section~\ref{sec:folds}: the labels-only student has no path for it, the mixed student has one, and the gap has the sign that leakage would produce. With one training run per target and three decimals on the leaderboard, a difference of 0.005 is consistent with zero, and it is also the size of the fold-0 difference between the two versions of the code, so we do not read it as evidence that report labels are the better target either. The mixture did not show a clear model-level advantage in this study. The only out-of-fold comparison with matched code was on fold~0, where the two targets lay within seed variation, and the post-deadline labels-only submission scored higher on the hidden test; the pooled out-of-fold comparison remains confounded by the change of code and by possible leakage. A clean answer would need a five-fold comparison with matched code and with teacher predictions made on the student's own folds, which we could not run.

\begin{table}[!t]
\caption{The same student trained on two targets, pooled out of fold and on the hidden test. Target: target agreement AUC on the 58 gold studies. The mixed student's out-of-fold scores may be optimistic (Section~\ref{sec:folds}) and come from the original code; the labels-only student and the leaderboard have no leakage path. The teacher was not submitted.}
\label{tab:labonly}
\small
\setlength{\tabcolsep}{7pt}
\begin{tabular}{lccccc}
\toprule
& & \multicolumn{2}{c}{Out of fold} & \multicolumn{2}{c}{Leaderboard} \\
\cmidrule(lr){3-4}\cmidrule(lr){5-6}
Model & Target & Expert labels & Report statements & Public & Private \\
\midrule
Student, report labels only & 0.892 & 0.857 & 0.859 & 0.891 & 0.877 \\
Student, 0.5 / 0.5 mixture & 0.927 & 0.876 & 0.863 & 0.887 & 0.872 \\
Teacher (public, other folds) & 0.892 & 0.896 & 0.879 & -- & -- \\
\bottomrule
\end{tabular}
\end{table}

\subsection{Explaining away helps a little out of fold and not visibly on the leaderboard}
\label{sec:eares}

\begin{figure}[!t]
\centering
\includegraphics[width=\linewidth]{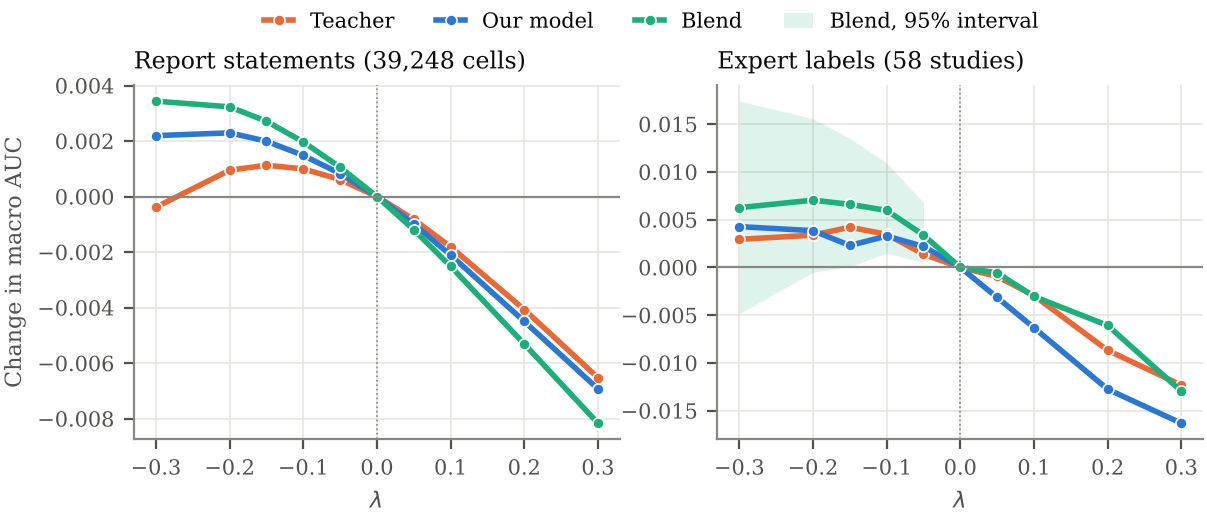}
\caption{Change in out-of-fold macro AUC under the rank correction of Equation~\eqref{eq:ea}. Negative $\lambda$ subtracts the ranks of co-occurring findings, positive $\lambda$ adds them. The band is the 95\% bootstrap interval of the blend.}
\label{fig:ea}
\end{figure}

Figure~\ref{fig:ea} shows Equation~\eqref{eq:ea} for the teacher, the student and their equal rank blend over $\lambda$ from $-0.3$ to $0.3$. Positive $\lambda$ lowered both references for all three score sets. Negative $\lambda$ raised the explicit AUC by up to 0.001 (teacher), 0.002 (student) and 0.003 (blend), and the gold AUC by 0.001 to 0.007. Appendix Table~\ref{tab:ea} gives all 15 negative settings with bootstrap intervals and permutation $p$ values. For the blend, the two smallest corrections ($\lambda=-0.05$ and $-0.1$) raised the gold AUC by 0.003 and 0.006, with Holm-adjusted $p$ of 0.015 for both; at $\lambda=-0.15$ the interval stays just above zero but the adjusted $p$ is 0.13. For the teacher or the student alone, no setting survives the adjustment (smallest adjusted $p$ 0.09). The DeLong-type test is less favourable to the blend: it gives $p$ 0.009 for both small settings and a Holm-adjusted $p$ of 0.13, so whether these two settings survive the adjustment depends on the test. These are gains of a few thousandths of macro AUC on 58 studies, while the explicit AUC moves by at most 0.003. In the blend at $\lambda=-0.1$, the explicit gain was largest for ACL (0.0044), the lateral meniscus (0.0040), PF OA (0.0036) and Baker's cyst (0.0028). The blend gained more than either model alone, which would fit the idea that averaging two models adds dependence between the findings; we did not test this further.

On the leaderboard, the correction was added to two public ensembles that both score 0.958 public: a fork of RSNA\_Versia\_7~\cite{versia2026notebook} and a fork of the CoAtNet, R6 and 2.5D MIL ensemble of goodpjw2008~\cite{goodpjw2026coatnet}. With $\lambda=-0.1$ on the first, the public score stayed at 0.958 and the private score went from 0.942 to 0.943. With $\lambda=-0.2$ on the second, the public score went from 0.958 to 0.957 and the private score stayed at 0.942 (Figure~\ref{fig:leaderboard}). Kaggle shows three decimals, so these changes are at the resolution of the leaderboard. We read the correction as a diagnostic of how much the scores of one finding depend on other findings. Its out-of-fold gain is small and rests on 58 studies, and the leaderboard did not show it, so we do not recommend it as a post-processing step.

\subsection{The two references can disagree}
\label{sec:refres}

\begin{figure}[!t]
\centering
\includegraphics[width=0.55\linewidth]{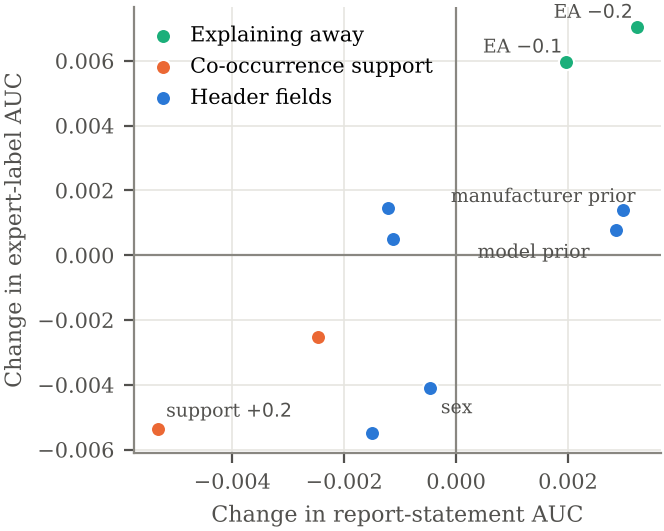}
\caption{Effect of each correction on the blend of student and teacher, on the report statements (horizontal) and on the expert labels (vertical). A correction that only moves right helps agreement with the reports and not with the images.}
\label{fig:references}
\end{figure}

Figure~\ref{fig:references} places every correction of the blend by its effect on the two references, and Table~\ref{tab:header} gives the numbers for all three score sets. Sex lowered the gold AUC by 0.004 to 0.005 and left the explicit AUC unchanged. Subtracting the mean rank of each scanner model lowered both references, by 0.038 to 0.049 gold at full strength, so the differences between scanner groups are mostly differences in case mix and should not be removed. The scanner prevalence prior is the instructive case. At weight 0.1 it raised the explicit AUC by 0.003 for all three score sets, as much as the best explaining-away setting, and changed the gold AUC by $+0.0002$ to $+0.0014$ for the manufacturer prior and $-0.003$ to $+0.002$ for the model prior. The explicit reference comes from reports written at the same sites, so a prior that encodes how often a site's radiologists write a finding can raise agreement with the reports without raising agreement with the images. A model selected on report statements alone would have kept this prior.

\begin{table}[!t]
\caption{Change in out-of-fold macro AUC on the report statements (Expl.) and the expert labels (Gold) for corrections built from DICOM headers and for co-occurrence support ($\lambda>0$, applied to all 12 findings).}
\label{tab:header}
\small
\setlength{\tabcolsep}{4pt}
\begin{tabular}{lcccccc}
\toprule
& \multicolumn{2}{c}{Teacher} & \multicolumn{2}{c}{Student} & \multicolumn{2}{c}{Blend} \\
\cmidrule(lr){2-3}\cmidrule(lr){4-5}\cmidrule(lr){6-7}
Correction & Expl. & Gold & Expl. & Gold & Expl. & Gold \\
\midrule
Sex, weight per finding (cross-fitted) & $+0.000$ & $-0.005$ & $-0.000$ & $-0.005$ & $-0.001$ & $-0.004$ \\
Scanner model centring, share 0.25 & $-0.002$ & $-0.006$ & $-0.001$ & $-0.008$ & $-0.002$ & $-0.006$ \\
Scanner model centring, share 1 & $-0.021$ & $-0.038$ & $-0.017$ & $-0.049$ & $-0.019$ & $-0.046$ \\
Manufacturer centring, share 0.25 & $-0.001$ & $-0.002$ & $-0.001$ & $-0.002$ & $-0.001$ & $+0.001$ \\
Field strength centring, share 0.25 & $-0.002$ & $+0.001$ & $-0.001$ & $-0.001$ & $-0.001$ & $+0.002$ \\
Manufacturer prevalence prior, 0.1 & $+0.003$ & $+0.001$ & $+0.003$ & $+0.000$ & $+0.003$ & $+0.001$ \\
Scanner model prevalence prior, 0.1 & $+0.003$ & $+0.002$ & $+0.003$ & $-0.003$ & $+0.003$ & $+0.001$ \\
Co-occurrence support, $\lambda=0.1$ & $-0.002$ & $-0.002$ & $-0.002$ & $-0.006$ & $-0.003$ & $-0.003$ \\
Co-occurrence support, $\lambda=0.2$ & $-0.004$ & $-0.008$ & $-0.005$ & $-0.012$ & $-0.005$ & $-0.005$ \\
\bottomrule
\end{tabular}
\end{table}

\subsection{Leaderboard}
\label{sec:lb}

Appendix~\ref{app:leaderboard} lists every scored submission. Our own model rose from 0.713 to 0.872 private between the first entry and the five-fold student, and the labels-only student scored 0.877 after the deadline. Blending our models into public ensembles changed their private scores by $-0.001$ to $+0.002$. Every submission lost 0.014 to 0.022 from public to private, about the same for our model and for the public ensembles. The best private score of the challenge was 0.960. Of the 6{,}299 teams, 1{,}297 scored exactly 0.942, the private score of the public ensembles we also forked, and our corrected fork shared 0.943 with 286 other teams (Appendix Figure~\ref{fig:lbdist}). Its displayed rank of 582 rests on the third decimal and on the order of submission among tied teams. It is the position shown on the Kaggle private leaderboard, not an award of the challenge.

\section{Limitations}
\label{sec:limits}

The expert labels cover 58 studies with 9 to 35 positives per finding, and the same studies were used to compare many variants. A bootstrap interval describes sampling noise under the observed sample and does not correct for the search; the Holm adjustment covers only the explaining-away family, and no adjustment was made across families. All gold results are exploratory, and an independent expert-labelled set would be needed to confirm any of them.

The report labels come from one language-model run on reports in nine languages, and we did not check its errors by language. The explicit reference inherits those errors, the omissions of the reports and the reporting habits of each site (Section~\ref{sec:refres}). Only adjudicated expert labels are released, so we cannot say how often two radiologists would disagree on these studies. Part of the gap between report labels and expert labels may lie within the range of ordinary reader disagreement, and for a report about that gap this is a real limit. Whether the expert readers saw the reports is not stated in the challenge material.

The teacher's folds differ from ours, so the mixed student's out-of-fold scores may be optimistic (Section~\ref{sec:folds}), and their advantage over a labels-only student did not appear on the one fold with matched code or on the test (Section~\ref{sec:labonly}). The target-agreement results, the labels-only student and the leaderboard scores are not affected.

An audit of the headers of all 24{,}371 training series (notebook~07) found the slice order sound: in every series the largest-spread axis is the series' own plane axis and the order equals the stack-normal order up to direction, with one direction per plane; 5.2\% of the series are tilted more than 20$^\circ$ from the nearest axis (at most 41$^\circ$), and row and column pixel spacing agree in every series. The mirroring is less clean. The side is read per series, from the DICOM tag in 49\% of the series and from the median $x$ position otherwise; 7.3\% of the series have neither and are left unmirrored, where both are present the position rule agrees with the tag in 94.5\% of the series, and in 3.9\% of the studies the series of one knee were given opposite sides. The decode functions come from the teacher's repository, so the teacher shares this. The target-level result does not depend on the affected gold studies (Table~\ref{tab:side}); the student's scores were not re-examined in this way. Deciding the side once per study and checking it across the series would avoid the problem. A failed slice is replaced by another slice without a count per study, and each slice has its own intensity window, which removes real intensity differences between neighbouring slices. A window of three cached slices covers a different physical thickness in each series. Besides the random crop, the only augmentation is an intensity gain and offset; rotations, elastic deformations and spacing-aware windows were not tried.

The reported student was trained once per fold with an unseeded data sampler. The fold-0 ablations are single runs on one fold, apart from three seeds of one configuration, and the cause of the difference between the original and the released code was not found. The window-thickness explanation for the menisci was not tested. The two five-fold students were each trained once, with different versions of the code, so we cannot judge whether their leaderboard difference exceeds run-to-run variation, and the labels-only score was obtained after the deadline.

The statistics of Table~\ref{tab:secondary} are descriptive: the sensitivity and specificity use a threshold set on the same 58 studies, and the calibration error is measured against report statements, not expert labels. The leaderboard reports three decimals, the public score was used for some choices, and our final rank comes from a public ensemble that we did not build and whose components we cannot redistribute. We report no calibration study at clinical operating points, no reader study and no external validation. The evidence is internal to the challenge data: it shows how report-derived and image-derived targets disagree with expert labels in this sample, and it says nothing about other sites, scanners or periods.

\section{Discussion}
\label{sec:discussion}

The clearest result concerns targets. A language-model reading of the report and an image model trained on it are about equally far from the expert labels, but for different findings, and an untuned equal mixture is closer to the experts than either. The report is nearly perfect for ligament tears, which radiologists name when they see them, better than the image model for PF OA, which it often misses but never calls present without the experts, and weak for effusion, Baker's cyst, contusion and synovitis, which the reports call absent, leave out or call present more often than the expert readers do (Table~\ref{tab:states}). These statements rest on 58 studies, and they concern targets. We did not observe that the better target made a better student: on the one fold with matched code the two targets lay within seed variation, and on the hidden test the labels-only student scored higher after the deadline (Section~\ref{sec:labonly}). For anyone training on report labels, the practical consequences are to measure the report labels per finding before using them, to compare trained models on data that no part of the target has seen, because an out-of-fold estimate can carry the target's own leaks, and to check any choice made on a report-derived reference against labels read from the images.

How knee MRI reports are written suggests why the two sources split along these lines, although we did not test it. Knee MRI is usually requested after a twisting injury or for pain, the referring question concerns the ligaments and menisci, and so a report states an ACL or MCL tear as present or absent almost every time (the ACL is unmentioned in 8\% of the reports, Figure~\ref{fig:coverage}); the cartilage of each compartment is usually graded as well. Effusion tends to be written only when the reader judges it more than physiological, and that threshold varies between readers. Synovitis is hard to call without contrast and is mentioned in 14\% of the reports. A small Baker's cyst is incidental and often left out. A bone bruise is often described as marrow oedema at a given site and not named as a contusion, so the reader of the report, human or language model, has to map the description to the finding. All four are visible on fluid-sensitive images, which is where an image model can supply what the text leaves out.

\subsection*{What the highest-ranked entries did}

After the challenge closed, eight of the ten highest-ranked teams (ranks 1 to 3, 5 to 8 and 10) published their solutions~\cite{kaggle2026first,kaggle2026second,kaggle2026third,kaggle2026fifth,kaggle2026sixth,kaggle2026seventh,kaggle2026eighth,kaggle2026tenth}. Table~\ref{tab:top} summarises how they built their training labels. All eight describe the labels as a main part of their work, all eight scored their label versions on the same 58 gold studies, and seven of them had a language model read the training images, with the report, before it or instead of it. Where they report numbers on the 58 studies, the reads that included the images scored highest: report-only reads reached 0.89 to 0.91 and reads that included the images 0.919 to 0.947 for the seventh-placed team~\cite{kaggle2026seventh}, and 0.899 against 0.942 for the second-placed team, with the largest gains for effusion (0.806 to 0.975) and Baker's cyst (0.803 to 0.979) and the report reading still ahead for PF OA~\cite{kaggle2026second}. The sixth-placed team, whose labels came from the reports and from image models trained on external expert grades, names this as its weakest point~\cite{kaggle2026sixth}. These write-ups describe different systems, all tuned on the same 58 studies, and none of them isolates the effect of reading the images, so we use them as context; they are not a controlled comparison and not an independent validation.

\begin{table}[!t]
\caption{Training labels of the published solutions among the ten highest-ranked teams, as described in their write-ups. Image read: a language model looked at the training images. Teacher: predictions of image models were mixed into the labels. External MRI: knees from other data sets (OAI, fastMRI) used for training or pretraining. --: not described.}
\label{tab:top}
\small
\setlength{\tabcolsep}{4pt}
\begin{tabular}{rc>{\raggedright\arraybackslash}p{6.4cm}ccl}
\toprule
Rank & Private & Label sources & Image read & Teacher & External MRI \\
\midrule
1 & 0.960 & agent reading images and report; region and lesion boxes & yes & -- & -- \\
2 & 0.960 & report reader, then images with report & yes & yes & OAI, fastMRI \\
3 & 0.959 & images with report, finding-specific rereads & yes & OAI cells & OAI \\
5 & 0.958 & calibrated report labels; images for three findings & yes & yes & OAI \\
6 & 0.958 & report reads and rules; image models trained on OAI grades & no & yes & OAI, fastMRI \\
7 & 0.957 & three reads: report with slices, images alone, verification & yes & yes & none \\
8 & 0.957 & images first, then report; per-finding image weights & yes & yes & none \\
10 & 0.957 & report reads; images with report; images alone & yes & yes & OAI \\
\midrule
582 & 0.943 & this work: report labels and an image teacher (own model 0.877) & no & yes & none \\
\bottomrule
\end{tabular}
\end{table}

Several details of these write-ups bear directly on our results. The eighth-placed team chose, per finding, how much weight the image reading gets against the report: most for fracture, synovitis and Baker's cyst, then effusion and contusion, least for PF OA, lateral OA and MCL~\cite{kaggle2026eighth}. This is the pattern of Figure~\ref{fig:complementarity}, found with other labels. Five teams mixed out-of-fold predictions into their labels, and where they give numbers the gains were modest: 0.947 to 0.952 on the 58 studies for the seventh-placed team, 0.931 to 0.935 for the eighth. The fifth-placed team describes the same leakage path as our Section~\ref{sec:folds}, treated its local scores as diagnostics for that reason, and found that folds sharing a hospital raised the AUC against report labels by 0.019 without raising it on the 58 expert-labelled studies~\cite{kaggle2026fifth}, the failure of the report reference we describe in Section~\ref{sec:refres}. The seventh-placed team found that synovitis was written according to hospital habit and corrected it per scanner model~\cite{kaggle2026seventh}, the habit that our scanner prior picked up. The tenth-placed team subtracted a share of the medial OA rank from the lateral and PF OA ranks and of the effusion rank from the synovitis rank, gained about 0.0004 out of fold and saw no change on the leaderboard~\cite{kaggle2026tenth}, much the fate of our explaining-away correction.

Two conclusions follow for our own work. Our image teacher read the images but learned from report labels, and our report labels never saw the images; the teams above had one reader look at both, and those reads gave their highest agreement with the expert labels. And because the mixed target of a report reader and an image model is what most of these teams also trained on, the result of Section~\ref{sec:labonly} suggests that such mixing should be judged on the test, or on folds that no teacher has seen, before it is credited with a gain.

The second result is a warning about references. The report statements are the only reference with tens of thousands of cells, and we used them for most training decisions, yet they rose for a prior that encodes site reporting rates and left the expert-label AUC where it was. When the reference is derived from reports, a change that improves it should also be checked against labels read from the images, however few they are, and a change that moves only the report reference should be treated as suspect.

The gap between target and model has a parallel in work on distillation. A more accurate teacher does not always give a better student~\cite{cho2019efficacy}, and students often fail to match their teacher's predictions even where they have the capacity to, with optimisation named as one reason~\cite{stanton2021does}. Our student was trained half on a teacher's soft predictions and stayed below that teacher on the expert labels (Section~\ref{sec:student}), and the target closer to the experts gave no better student (Section~\ref{sec:labonly}). Both fit those findings, although our setting differs from theirs: the teacher is an image model with another input format, read from fixed out-of-fold files, and the mixed target also carries the report labels.

The student did not reach its teacher, and on the leaderboard it was 0.07 below the public ensembles, which use CoAtNet and ConvNeXt models trained with OAI knees. The explaining-away correction gave small gains out of fold, which for the blend survive Holm adjustment within the 15 settings under the permutation test but not under the DeLong-type test, and no effect on the leaderboard beyond its display resolution.

\subsection*{Implications for challenge design}

Several limits of this report come from what the challenge released, and some would be cheap to remove in a future challenge. Releasing the label of each expert reader along with the adjudicated label would show how far report-derived labels are from the experts compared with how far the experts are from each other. Shared out-of-fold predictions are safe to train on only when their folds are published with them; the public teacher we used came without its folds, and the leakage path of Section~\ref{sec:folds} followed. A label file that keeps the state of each finding lets users separate a finding the report leaves out from one it negates, and Table~\ref{tab:states} shows that these are different errors for different findings. Every team tuned its labels on the same 58 studies, so a second expert-labelled set held back until the end, or an expert set large enough to split, would let teams check label choices on studies they had not used for selection; with 9 MCL and 11 lateral OA positives, the present set cannot rank close variants for those findings.

\section{Code, data and licences}
\label{sec:availability}

The code, the frozen predictions, the logs and the data tables described here are at \url{https://github.com/soheilkooklan/rsna-knee-report-labels} (release v1.0). One command recomputes every number of Sections~\ref{sec:targetres} to~\ref{sec:refres} and draws the figures from the released files: the report labels and teacher predictions (MIT, from NTejas-1~\cite{ntejas2026repo}), our out-of-fold predictions, the fold manifest, one DICOM header per training study and the header audit of every training series. The release reproduces the analyses; it does not reproduce the leaderboard. It does not retrain the student, it holds no trained checkpoints, and it cannot check the hidden leaderboard scores, which are recorded as Kaggle showed them. Unit tests cover the target construction, the slice order, the rank correction, checkpoint discovery and the statistics, and a CPU test runs the training and submission notebooks end to end on synthetic DICOM data inside a temporary folder; the log of a clean run is included. The notebooks that trained the student and wrote the submissions are generated by one script, and their defaults are the reported run. The reported folds were trained before the per-sample random streams of the released code were added, so a retraining gives close but not identical weights; the per-epoch validation scores of the reported runs, transcribed from the Kaggle logs, are released as a table. The later runs (Table~\ref{tab:ablation} and the labels-only student) were made with the notebook cells of the earlier version, which did not write per-run manifests. A table in the release records their settings, seeds and checkpoint hashes, and their Kaggle logs are included, but their package versions were not recorded.

The RadImageNet ResNet50 weights~\cite{mei2022radimagenet} were read from the file \texttt{ResNet50.pt} of the Kaggle dataset \texttt{marwanmath/resnet-50-radimagenet-marwan}~\cite{marwan2026radimagenet}. The hash of that file was not recorded for the reported runs (the released training code now logs it); the weights are not part of the release, and the original weights are distributed by the RadImageNet authors under their own terms. Competition images and reports are not redistributed. The public ensembles in rows 4 to 13 of Table~\ref{tab:subs} are not part of the release. The submission that set our rank is the newer version of RSNA\_Versia\_7, unchanged, with one added cell that applies Equation~\eqref{eq:ea} with $\lambda=-0.1$ to its \texttt{submission.csv}; that cell is in the release. We did not access OAI data; the public models in those ensembles were trained by their authors with OAI knees.

\section*{Ethics, funding and competing interests}

This is a secondary analysis of de-identified data that RSNA released for the challenge, used under the challenge rules; no new data were collected, and no separate ethics review was sought. The work received no specific funding. The author declares no competing interests.

\section*{Acknowledgements}
We thank RSNA, the annotating radiologists and Kaggle for the data and the challenge. The report labels and the teacher predictions come from the NTejas-1 repository~\cite{ntejas2026repo}. The public ensembles in our final submissions are the work of their authors: RSNA\_Versia\_7 by evgendvorkin~\cite{versia2026notebook}, the CoAtNet ensemble by goodpjw2008~\cite{goodpjw2026coatnet}, which builds on the CoAtNet checkpoints of nartaa~\cite{nartaa2026coatnet} and the compact models and per-finding weights of heliosli~\cite{heliosli2026blend}, and the Speedy Raptors notebook. Their authors ask users to include the following acknowledgement for the OAI-trained weights: Data and/or research tools used in the preparation of this manuscript were obtained and analyzed from the controlled access datasets distributed from the Osteoarthritis Initiative (OAI), a data repository housed within the NIMH Data Archive (NDA). OAI is a collaborative informatics system created by the National Institute of Mental Health and the National Institute of Arthritis, Musculoskeletal and Skin Diseases (NIAMS) to provide a worldwide resource to quicken the pace of biomarker identification, scientific investigation and OA drug development. Dataset identifier(s): 10.15154/0hcg-f676~\cite{oai2026}. AI assistants were used in this work: Claude (Anthropic) debug code, edit the text, and ChatGPT (OpenAI) and Grok (xAI) to review the paper and the code. The author directed the work, ran the GPU training on Kaggle, chose the submitted configurations, checked the numbers against the logs and takes full responsibility for the content.

\bibliographystyle{unsrtnat}
\bibliography{refs}

\appendix
\section{Settings}
\label{app:settings}

\begin{table}[!htbp]
\caption{Initial values of the slot prior, a learnable bias on the finding-to-slot attention logits. Slots: sagittal fluid-sensitive, sagittal other, coronal fluid-sensitive, coronal other, axial fluid-sensitive, axial other.}
\label{tab:prior}
\small
\begin{tabular}{lcccccc}
\toprule
Finding & Sag.\ fluid & Sag.\ other & Cor.\ fluid & Cor.\ other & Ax.\ fluid & Ax.\ other \\
\midrule
ACL & 1 & 1 & 0.5 & 0 & 0 & 0 \\
MCL & 0 & 0 & 1 & 0.5 & 0.5 & 0 \\
Medial and lateral meniscus & 1 & 1 & 1 & 0.5 & 0 & 0 \\
Medial and lateral OA & 0.5 & 0.5 & 1 & 1 & 0 & 0 \\
PF OA & 0.5 & 0 & 0 & 0 & 1 & 1 \\
Effusion, synovitis, Baker's cyst & 1 & 0 & 0 & 0 & 1 & 0 \\
Contusion & 1 & 0 & 1 & 0 & 0.5 & 0 \\
Fracture & 1 & 0.5 & 1 & 0.5 & 0 & 0 \\
\bottomrule
\end{tabular}
\end{table}

After training, the prior stays close to its initial values (mean absolute change 0.11, largest 0.43, for PF OA on the axial series without fat suppression). Recorded on the held-out studies of each fold (notebook~08), the attention of each finding is spread over the series: no finding gives more than 0.31 of its attention to one slot on average, the axial slot without fat suppression, often empty, receives 0.03, and the five folds agree within 0.02. The prior is therefore a mild bias, and the attention weights do not show which series a finding depends on in any strong sense.

\begin{table}[!htbp]
\caption{Settings of the reported student (family \texttt{r50v3}).}
\label{tab:impl}
\small
\begin{tabular}{>{\raggedright\arraybackslash}p{3.4cm}>{\raggedright\arraybackslash}p{11.8cm}}
\toprule
Part & Setting \\
\midrule
Cache & 4{,}407 $\times$ 6 $\times$ 16 $\times$ 256 $\times$ 256 bytes in three shards built on CPU; 6 to 94\% band; 140\,mm crop; deepest series per slot; slices sorted along the largest-spread patient axis \\
Instance & 3 consecutive cached slices as channels; 4 random windows per slot in training, all 14 at test \\
Encoder & RadImageNet ResNet50, global average pooling, 2{,}048 features, input in $[-1,1]$ \\
Slot pooling & gated attention, hidden size 256 \\
Finding head & LayerNorm, linear to 512, GELU, learnable slot embedding, one query per finding, slot prior, dropout 0.2 \\
Augmentation & random 224\,px crop of 256\,px, gain $[0.9,1.1]$, offset $[-8,8]$; no flips \\
Optimiser & AdamW, encoder $2\cdot10^{-4}$, head $2\cdot10^{-3}$, weight decay 0.01, warm-up 8\% then cosine, 8 epochs \\
Batch & 2 studies per step on two T4 GPUs (DataParallel), accumulation 4, mixed precision, clipping at 3 \\
Weight averaging & mean of the weights after epochs 5, 6 and 7 \\
Targets & 0.5 report labels + 0.5 teacher; gold studies at weight 2 \\
Test & fold models averaged in probability space; families combined in rank space; test cache rebuilt from the hidden DICOM inside the notebook \\
\bottomrule
\end{tabular}
\end{table}

\section{Secondary metrics, calibration and the explaining-away grid}
\label{app:secondary}

\begin{table}[!htbp]
\caption{Descriptive secondary metrics, out of fold. AP: average precision against the expert labels (gold prevalence in brackets). Sens./Spec.: sensitivity and specificity of the student on the 58 gold studies at a threshold set on the same studies so that it flags as many studies as the gold prevalence; this is a description of ranking, not an operating point. ECE: expected calibration error of the student against the stated report cells (10 bins; stated cells per finding in Figure~\ref{fig:reliability}), not against the expert labels.}
\label{tab:secondary}
\small
\setlength{\tabcolsep}{6pt}
\begin{tabular}{lccccc}
\toprule
Finding & AP student & AP teacher & Sens. / Spec. & ECE \\
\midrule
ACL (0.41) & 0.867 & 0.899 & 0.79 / 0.85 & 0.03 \\
MCL (0.16) & 0.667 & 0.546 & 0.56 / 0.92 & 0.05 \\
Medial meniscus (0.45) & 0.835 & 0.918 & 0.81 / 0.84 & 0.03 \\
Lateral meniscus (0.40) & 0.738 & 0.781 & 0.65 / 0.77 & 0.04 \\
Medial OA (0.26) & 0.919 & 0.901 & 0.80 / 0.93 & 0.03 \\
Lateral OA (0.19) & 0.745 & 0.648 & 0.55 / 0.89 & 0.04 \\
PF OA (0.36) & 0.793 & 0.805 & 0.67 / 0.81 & 0.03 \\
Effusion (0.60) & 0.985 & 0.971 & 0.89 / 0.83 & 0.05 \\
Synovitis (0.47) & 0.801 & 0.865 & 0.67 / 0.71 & 0.33 \\
Baker's cyst (0.21) & 0.838 & 0.859 & 0.75 / 0.94 & 0.03 \\
Contusion (0.33) & 0.896 & 0.894 & 0.79 / 0.90 & 0.07 \\
Fracture (0.31) & 0.849 & 0.881 & 0.78 / 0.90 & 0.03 \\
\midrule
Macro & 0.828 & 0.831 & 0.72 / 0.86 & 0.06 \\
\bottomrule
\end{tabular}
\end{table}

\begin{figure}[!htbp]
\centering
\includegraphics[width=\linewidth]{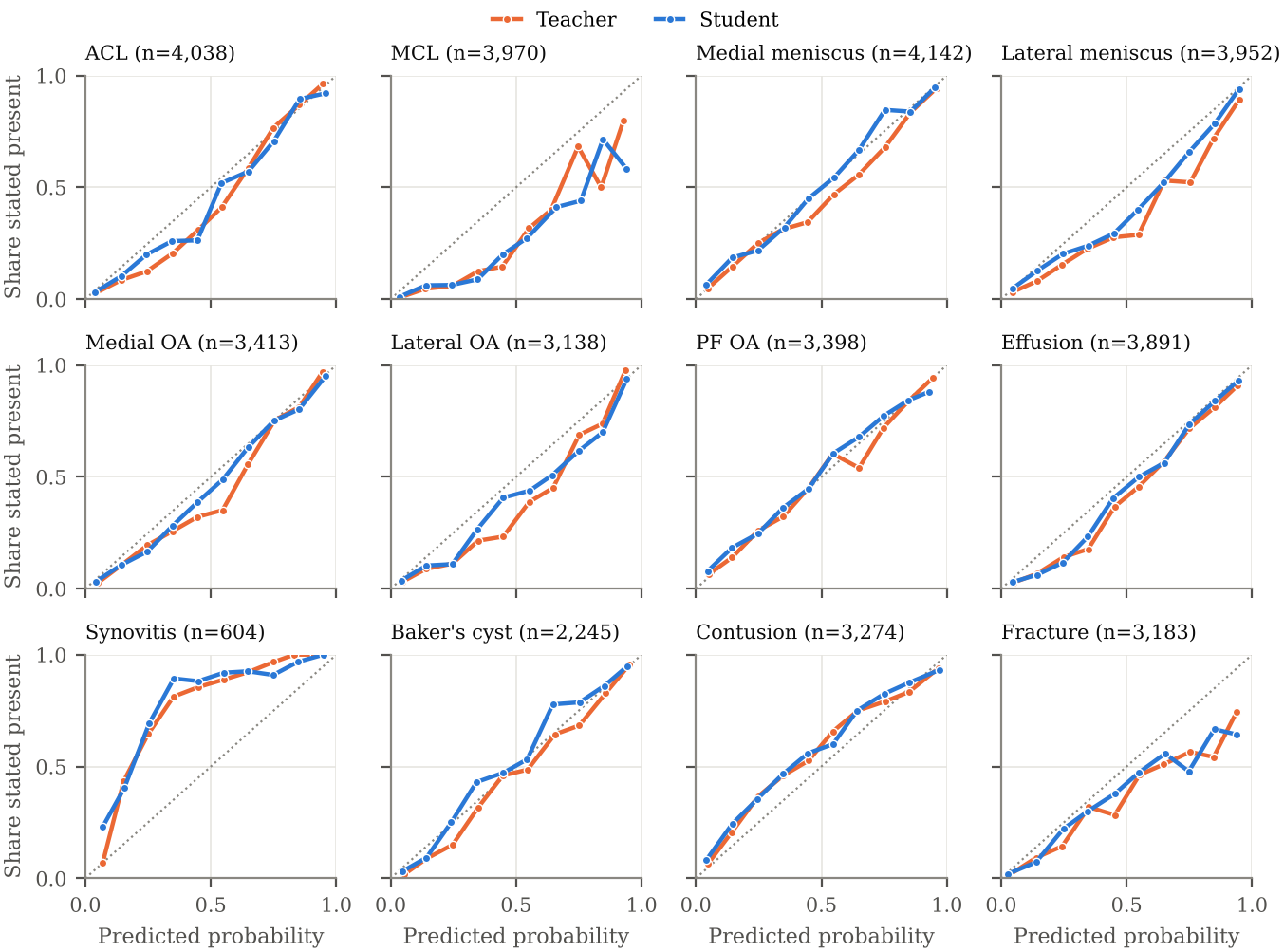}
\caption{Reliability of the out-of-fold probabilities of the student and the teacher against the stated report cells ($n$ per finding), ten equal-width bins. The reference is what the reports state, not the expert labels.}
\label{fig:reliability}
\end{figure}

The calibration error of synovitis (0.33) comes from the reference more than from the model. Of the 604 reports that mention synovitis, 82\% call it present, because it is written mostly when seen; the student, trained on all studies, gives these studies probabilities that fall well below that share (Figure~\ref{fig:reliability}). For the other findings the curves of student and teacher lie close to the diagonal.

\begin{table}[!htbp]
\caption{Explaining-away correction on out-of-fold scores. Gold: observed difference in macro AUC to $\lambda = 0$ on the 58 gold studies, with 95\% bootstrap interval; $p$: paired permutation test; DeLong: paired test of DeLong type (Section~\ref{sec:stats}); Holm: adjusted over the 15 rows, for each test separately.}
\label{tab:ea}
\small
\setlength{\tabcolsep}{4pt}
\begin{tabular}{lcccccccc}
\toprule
& & & & \multicolumn{2}{c}{Permutation} & \multicolumn{2}{c}{DeLong type} \\
\cmidrule(lr){5-6}\cmidrule(lr){7-8}
Scores & $\lambda$ & Explicit & Gold difference (interval) & $p$ & Holm $p$ & $p$ & Holm $p$ \\
\midrule
Teacher & $-0.05$ & 0.880 & $+0.0014$ ($-0.001$, $0.004$) & 0.054 & 0.539 & 0.242 & 1.000 \\
 & $-0.10$ & 0.880 & $+0.0035$ ($-0.000$, $0.007$) & 0.007 & 0.091 & 0.038 & 0.493 \\
 & $-0.15$ & 0.880 & $+0.0042$ ($-0.002$, $0.010$) & 0.056 & 0.539 & 0.121 & 1.000 \\
 & $-0.20$ & 0.880 & $+0.0034$ ($-0.004$, $0.011$) & 0.262 & 1.000 & 0.335 & 1.000 \\
 & $-0.30$ & 0.879 & $+0.0029$ ($-0.006$, $0.013$) & 0.449 & 1.000 & 0.514 & 1.000 \\
Student & $-0.05$ & 0.864 & $+0.0022$ ($-0.002$, $0.007$) & 0.226 & 1.000 & 0.303 & 1.000 \\
 & $-0.10$ & 0.864 & $+0.0033$ ($-0.002$, $0.010$) & 0.096 & 0.767 & 0.183 & 1.000 \\
 & $-0.15$ & 0.865 & $+0.0023$ ($-0.004$, $0.010$) & 0.410 & 1.000 & 0.480 & 1.000 \\
 & $-0.20$ & 0.865 & $+0.0038$ ($-0.004$, $0.013$) & 0.263 & 1.000 & 0.357 & 1.000 \\
 & $-0.30$ & 0.865 & $+0.0043$ ($-0.007$, $0.017$) & 0.379 & 1.000 & 0.458 & 1.000 \\
Blend & $-0.05$ & 0.880 & $+0.0034$ ($0.000$, $0.007$) & $<$0.001 & 0.015 & 0.009 & 0.132 \\
 & $-0.10$ & 0.881 & $+0.0060$ ($0.002$, $0.011$) & $<$0.001 & 0.015 & 0.009 & 0.132 \\
 & $-0.15$ & 0.882 & $+0.0066$ ($0.000$, $0.014$) & 0.011 & 0.132 & 0.040 & 0.493 \\
 & $-0.20$ & 0.882 & $+0.0070$ ($-0.001$, $0.016$) & 0.039 & 0.429 & 0.068 & 0.750 \\
 & $-0.30$ & 0.882 & $+0.0063$ ($-0.005$, $0.017$) & 0.197 & 1.000 & 0.242 & 1.000 \\
\bottomrule
\end{tabular}

\smallskip
\footnotesize At $\lambda=0$ the explicit AUC is 0.879 (teacher), 0.863 (student) and 0.879 (blend).
\end{table}

\begin{figure}[!htbp]
\centering
\includegraphics[width=\linewidth]{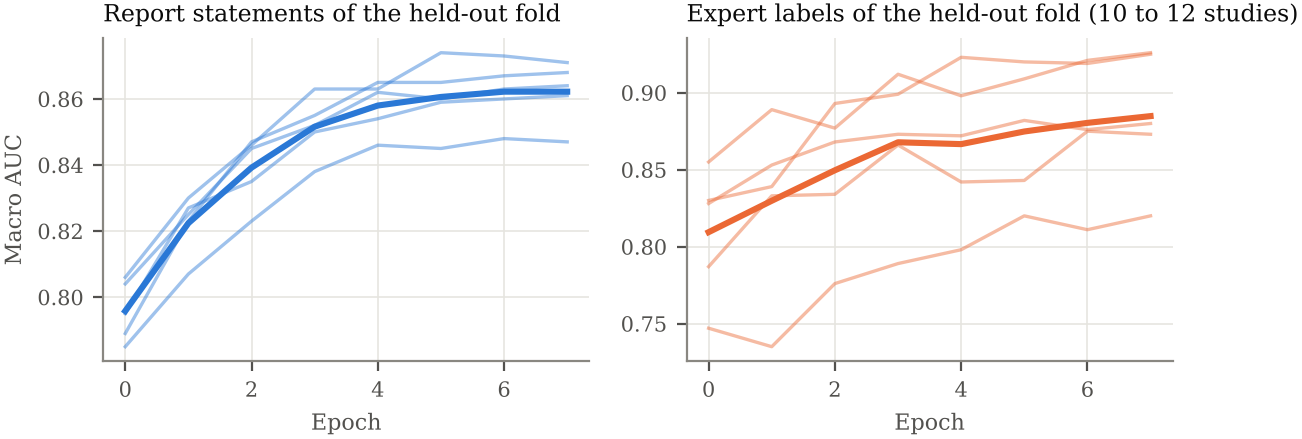}
\caption{Validation macro AUC per epoch for the five folds of the student (thin) and their mean (thick).}
\label{fig:epochs}
\end{figure}

\section{Leaderboard record}
\label{app:leaderboard}

\begin{figure}[!htbp]
\centering
\includegraphics[width=0.5\linewidth]{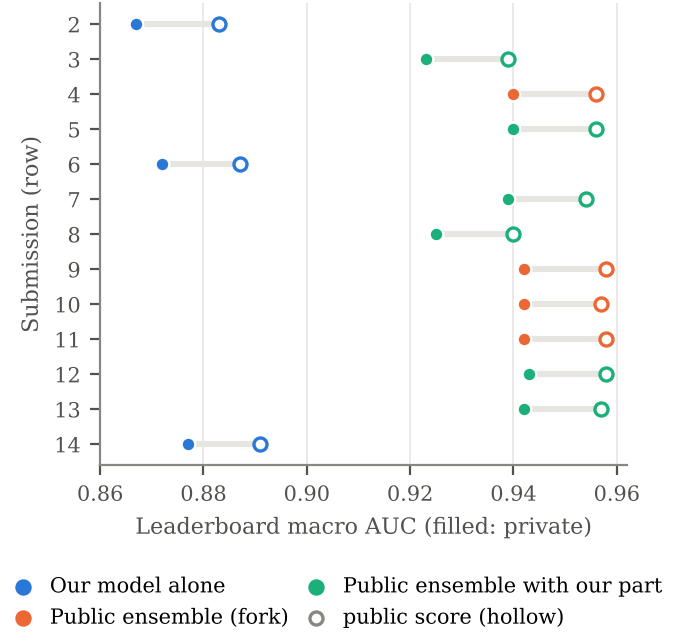}
\caption{Public (hollow) and private (filled) score of every scored submission after the first entry (0.735 public, 0.713 private), numbered as in Table~\ref{tab:subs}.}
\label{fig:leaderboard}
\end{figure}

\begin{figure}[!htbp]
\centering
\includegraphics[width=0.85\linewidth]{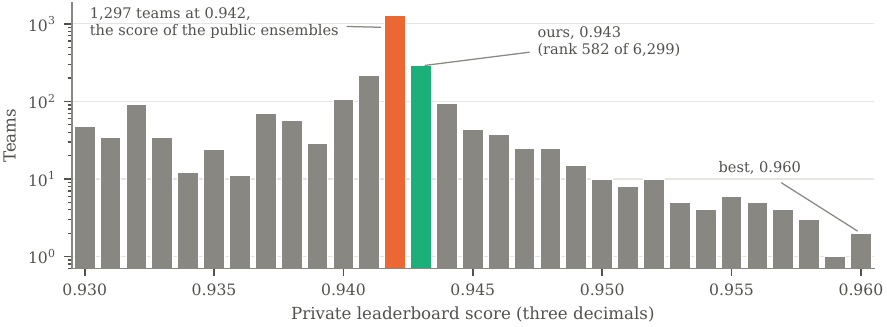}
\caption{Teams per private score between 0.930 and 0.960 (log scale). 2{,}240 teams scored below 0.900 and are not shown; our own models (0.872 and 0.877) are among them.}
\label{fig:lbdist}
\end{figure}

Table~\ref{tab:subs} and Figure~\ref{fig:leaderboard} give every scored submission. Blended into the Versia ensemble at weights 0.1 and 0.2, our models changed its private score by 0.000 and $-0.001$; in a blend with the public Speedy Raptors notebook, replacing our fold-0 model by the five-fold one raised the private score by 0.002. A blend of two full public ensembles in one notebook ran past the nine-hour limit on the hidden test set, as the decoding time of about 40\,s per study in the public run suggested. Row 14 was submitted after the deadline and is not a challenge entry. Figure~\ref{fig:lbdist} shows where the 6{,}299 teams ended.

\begin{table}[!htbp]
\caption{Scored submissions in the order they were made. Weights are rank-blend weights of our model. EA: explaining-away correction. Private scores were seen only after the challenge closed.}
\label{tab:subs}
\small
\begin{tabular}{r>{\raggedright\arraybackslash}p{8.6cm}cc}
\toprule
& Submission & Public & Private \\
\midrule
1 & First entry & 0.735 & 0.713 \\
2 & Student on the earlier cache, fold 0 only & 0.883 & 0.867 \\
3 & Speedy Raptors notebook + row 2 at weight 0.3 & 0.939 & 0.923 \\
4 & RSNA\_Versia\_7, older version~\cite{versia2026notebook} & 0.956 & 0.940 \\
5 & Row 4 + row 2 at weight 0.1 & 0.956 & 0.940 \\
6 & Student, five folds & 0.887 & 0.872 \\
7 & Row 4 + row 6 at weight 0.2 & 0.954 & 0.939 \\
8 & Speedy Raptors notebook + row 6 at weight 0.3 & 0.940 & 0.925 \\
9 & RSNA\_Versia\_7, newer version & 0.958 & 0.942 \\
10 & heliosli ensemble, ``best public dataset, 0.9586'' & 0.957 & 0.942 \\
11 & goodpjw2008 CoAtNet + R6 + 2.5D MIL~\cite{goodpjw2026coatnet} & 0.958 & 0.942 \\
12 & Row 9 + EA, $\lambda=-0.1$ (final rank) & 0.958 & 0.943 \\
13 & Row 11 + EA, $\lambda=-0.2$ & 0.957 & 0.942 \\
14 & Student, report labels only, five folds (after the deadline) & 0.891 & 0.877 \\
\bottomrule
\end{tabular}
\end{table}

\end{document}